\documentclass[a4paper,10pt]{article}

\usepackage{amssymb}

\usepackage[figuresright]{rotating}

\usepackage[section]{algorithm}
\floatname{algorithm}{Algorithme}
\usepackage{algorithmicx}

\usepackage{amsmath}
\usepackage{ragged2e}

\newenvironment{keywords}{\textbf{Keywords : }}{}

\title{Anopheles number prediction on environmental and climate variables using Lasso and stratified two levels cross validation}
\author{ Bienvenue Kouway\`e$^{1,\; 2, \; 3, \;  *}$}
\date{ 1- Universit\'e d'Abomey-Calavi, International Chair in Mathmatical Physic and Applications (ICMP:UNESCO-Chair), Abomey-Calavi, B\'enin\\ 
 2- Universit\'e Paris 1  Panth\'eon Sorbonne,  Laboratoire SAMM, Paris, France.\\
 3- Laboratoire SAMM Paris 1 France\\
 * E-mail :   kouwaye2000@yahoo.fr
 }
 
\begin{document}

\maketitle
 \justifying
\section*{Abstract} 
This paper deals with prediction of anopheles number using environmental and climate variables. The variables selection is performed 
by an automatic machine learning  method 
 based on Lasso and stratified two levels cross validation. Selected variables are debiased   while the prediction
is generated by simple GLM (Generalized linear model). Finally, the results reveal to be qualitatively better, at selection, the prediction,
and the CPU time point of view than those obtained by B-GLM method.\\
\begin{keywords}
 Malaria, variables selection, Lasso, cross validation, prediction
\end{keywords}
\section*{Introduction}
Malaria  is endemic in developing countries, mainly in sub-Saharan Africa. 
Among parasitic diseases 
,  malaria  is the main cause of mortality  especially for children under five years
of age  in Africa \cite{WHO}.
Generally, cohort studies take place in endemic areas. some of them are  for  characterizing  the malaria risk (the number of anopheles caught). 
These cohorts  studies are on newborn babies and pregnant women.
They are  introduced    to know about the immunity of newborn face to malaria  and the setting
 of this immunity. They also 
help to know  the determinants implicated in the  appearance 
of first malaria infections on the newborn,  the infant exposure to malaria and the malaria risk.
It is important to know 
the interaction between the host and the parasite, the repartition of malaria  
risk at small scale. This repartition  and the  malaria risk exposure present simultaneously spatial and temporal 
dependencies and non-homogeneous at small scale (house level) \cite{cottrell2012malaria}. 
Recent studies 
highlight 
 in the repartition of the malaria risk and transmission, 
 vector profile, 
 ecology, 
 seasonality, 
  characteristics
of habitats and the inhabitants practices \cite{cottrell2012malaria, dery2010patterns, craig1999climate, gu2005habitat}. 
It is necessary to understand 
the relation  among 
 the malaria risk and the environmental and climate factors. \\
In this study, we propose an automatic  machine learning method for variables selection  combining Lasso, GLM and two
levels cross validation in epidemiology
context.  One of the aim of such  approach is to   overcome  the pre-treatments of experts in medicine and epidemiology
on collected data.
The  proposed approach 
 uses every 
  available explanatory variables without  treatment and 
generates 
 automatically   all the  
 interactions  among 
 them. This leads to high dimension of variable selection. 
 The Lasso method proposed by Tibshirani \cite{tibshirani1996regression} is a regularized estimation 
approach for regression model using an $L_{1}$-norm and constraining the regression coefficients.
 The results of this method  is that all coefficients 
 are shrunken 
 toward zero and some are set exactly to zero. This method simultaneously  performs selection  and estimation, 
 and it is robust for variables selection in high dimension. 
 In some cohort studies, the number of observations is lower. The classical re-sampling method used is cross validation.  
 It is also well known that cross validation may lead to over fitting and one alternative solution 
is $percentile-cv$ \cite{Ng97preventing"overfitting"}. To avoid that 
 in learning stage, we propose 
 a stratified cross validation  with two levels.  According to the nature of the target variable, family of models 
 used for features selection, estimation and prediction are generally
linear models, generalized models, mixed models, generalized mixed models, multilevel modeling \cite{cottrell2012malaria}.
The target variable is the number of Anopheles, the main  characteristic of malaria risk, 
 it is an account variable. It is well
known that the Lasso coefficients are biased. 
A combination of GLM and Lasso (GLM-Lasso) is performed based on a 
cross validation with two levels, and a simple GLM is used
to debiased Lasso coefficients because of the family of the target variable.
For malaria risk prediction, four strategies of variables
selection based on GLM-Lasso and cross validation :
LDLM, LDLS, FVM, FVS are implemented. These strategies use some criteria : 
the mean, the quadratic risk, the absolute risk of the predictions,
and the deviance of the model. Each strategy is applied on four groups of covariables (original, original with village, recoded, recoded with village).
Most of the algorithms implemented in our work are based on \cite{Friedman2015glmnet, goeman1, zou2005regularization}.
The results are compared to those obtained by reference method 
(B-GLM) which uses a backward procedure combine with a GLM \cite{cottrell2012malaria}.

The results obtained by such 
 procedure are clearly better improved compared
to those obtained by the B-GLM method taken as the reference method. The improvement is about 
 all properties 
such as the quality of the selection and prediction. Moreover, 
the CPU time used to display
our program is smaller than the one 
required by the reference method and only few climate and environmental variables are the main 
factors associated to the malaria risk exposure with an improved accuracy.
\section*{ Materials and Methods}
\subsection*{Materials}
In this section, we briefly recall the description of the study area, the mosquito collection  
 and identification as well as the data,  
  and 
related variables. For more details, see \cite{cottrell2012malaria}.

\subsubsection*{Study area }
The study was conducted in the district of Tori-Bossito
(Republic of Benin), from July 2007 to July 2009. Tori-Bossito is on the coastal plain of Southern Benin, 40 kilometers
north-east of Cotonou. This area has a subtropical climate and
during the study,  
 the rainy season lasted from May to October.
Average monthly temperatures varied between 27$^\circ$C and 31$^\circ$C.
The original equatorial forest has been cleared and the vegetation
is characterized by bushes with sparse trees, a few oil palm
plantations, 
 and farms. The study area contained nine villages
(Avam\'e centre, Gb\'edjougo, Houngo, Anavi\'e, Dohinoko, Gb\'etaga,
Tori Cada Centre, Z\'eb\`e, 
 and Zoungoudo). Tori Bossito was
recently classified as mesoendemic with a clinical malaria
incidence of about 1.5 episodes per child per year \cite{damien2010malaria}.
Pyrethroid-resistant malaria vectors are present \cite{Djenontin}.

\subsubsection*{Mosquito collection and identfication}
Entomological surveys based on human landing catches (HLC)
were performed in the nine villages every six weeks for two years
(July 2007 to July 2009). Mosquitoes were collected at four catch
houses in each village over three successive nights (four indoors and
four outdoors, i.e. a total of 216 nights every six weeks in the nine
villages). Five catch sites had to be changed in the course of the study
(2 in Gbedjougo, 1 in Avam\`e, 1 in Cada, 1 in Dohinoko) and a total
of 19 data collections were performed in the field from July 2007
to July 2009. In total, data from 41 catch sites are available.
Each collector caught of predictional 
 mosquitoes landing on the lower legs
and feet between 10 pm and 6 am. 
All 
 mosquitoes were held in
bags labeled with the time of collection. The following morning,
mosquitoes were identified on the basis of morphological criteria
\cite{Gillies1,Gillies2}. All An. gambiae complex and An. funestus mosquitoes were
stored in individual tube 
 with silica gel and preserved at 220$^\circ$C.
P. falciparum infection rates were then determined on the head and
thorax of individual anopheline specimens by CSP-ELISA \cite{Wirtz}.
\subsubsection*{Environnement and behavioral data}
Rainfall was recorded twice a day with a pluviometer in each
village. In and around each catch site, the following information
was systematically collected: (1) type of soil (dry lateritic or humid
hydromorphic)—assessed using a soil map of the area (map IGN
Benin at 1/200 000 e , sheets NB-31-XIV and NB-31-XV, 1968)
that was georeferenced and input into a GIS; (2) presence of areas
where building constructions are ongoing with tools or holes
representing potential breeding habitats for anopheles; (3)
presence of abandoned objects (or ustensils) susceptible to be used
as oviposition sites for female mosquitoes; (4) a watercourse
nearby; (5) number of windows and doors; (6) type of roof (straw or
metal); (7) number of inhabitants; (8) ownership of a bed-net or (9)
insect repellent; and (10) normalized difference vegetation index
(NDVI) which was estimated for 100 meters around the catch site
with a SPOT 5 High Resolution (10 m colors) satellite image
(Image Spot5, CNES, 2003, distribution SpotImage S.A) with
assessment of the chlorophyll density of each pixel of the image.
Due to logistical problems, rainfall measurements are only
available after the second entomological survey. Consequently, we
excluded the first and second survey 
 (performed in July and
August 2007 respectively) from the statistical analyses.
\subsubsection*{Variables}
The dependent variable was the number of
Anopheles collected in a house over the three nights of each
catch and the explanatory variables were the environmental
factors, i.e. the mean rainfall between two catches (classified
according to quartile), the number of rainy days in the ten days
before the catch (3 classes [0–1], [2–4], $>$4 days), the season
during which the catch was carried out (4 classes: end of the dry
season  from February to April; beginning of the rainy season from May to
July; end of the rainy season from August to October; beginning of the
dry season from November to January), the type of soil 100 meters
around the house (dry or humid), the presence of constructions
within 100 meters of the house (yes/no), the presence of
abandoned tools within 100 meters of the house (yes/no), the
presence of a watercourse within 500 meters of the house (yes/no),
NDVI 100 meters around the house (classified according to
quartile), the type of roof (straw or Sheet metal), the number of
windows (classified according to quartile), the ownership of bed
nets (yes/no), the use of insect repellent (yes/no), 
 and the number
of inhabitants in the house (classified according to quartile). These pre-treatments 
  based
on the knowledge of experts in entomology and medicine operated 
on some  original variables  generate  a  second type of  covariables called recoded variables. 
The  Original and recoded variables are described in Tables
\ref{tab:Originales variables} and \ref{tab:Recoded variables}. 
Two  types of covariables set are used :  the first set, the original covariables with all covariables obtained by interactions,  
the second set, the recoded covariables with all covariables obtained by interactions. For
knowing the effect of the village on the selection method and prediction, four groups of covariables are considered : Group 1 (original variables), 
Group 2 (original variables with village as fixed effect),
Group 3 (recoded variables) and Gropu 4 (recoded variables with village as fixed effect)
\subsubsection*{Interactions between variables}
\label{Interactions}
Generally,  experts in epidemiology and medicine  choose some interactions according to their knowledge and experience.
 To avoid this way of making, we generate automatically all interactions in the full set of explanatory variables 
 used in the  
  model. This implies that the number of variables exponentially grows  and the classical method of variable selection 
  fails. The algorithm developed automatically  learns with all variables and all interactions and provides the optimal set of variables
for prediction. Assume that $p$  is the number of original covariables, the number of covariables including interactions
  is $N_{cov}$,  $\mathcal{V_O}= (V_1, \ldots, V_p)$  is a vector of  original covariables. The set of interactions covariables
 is defined as $\mathcal{I}_{\mathcal{V_O}} = \{V_i: V_j, 1\leq i,j\leq p,\;\; i\neq j \}$, interactions available for : 
 numerical crossed with numerical, numerical crossed with non-numerical  and non-numerical crossed with non-numerical covariables.
 Therefore, the number of covariables of interaction is   $p(p-1)/2$ and the total
 number of variables is 
 $N_{cov}=p+p(p-1)/2$. Assume that the number of observations is $N_{obs}$.
 \begin{enumerate}
  \item  \textbf{ Numerical variable crossed with  numerical variable} :\\
  $V_k$ and $V_l$ are two numerical variables. The variable of interaction obtained from  $V_k$ and $V_l$ is noted 
 $V_k : V_l$ and defined as :
 \begin{equation*}
  (V_k : V_l)_i = (V_k)_i \times (V_l)_i , \; 1 \leq i \leq N_{obs}
 \end{equation*}
 \item \textbf{Numerical variable crossed with  non-numerical variable} :\\
  $V_k$ is a numerical variable  and $V_l$ a non-numerical variable with $d_l$ modalities.
  $V_l$ is considered as a numerical variable with $d_l$-dimension. It can be replaced by 
  the indicators of its modalities. Suppose that the modalities are $V_{lq}, \; \; 1 \leq q \leq d_l$.
  The indicator  $I_{lq}$ associated to $V_{lq}$ is defined as :
  \begin{displaymath}     
   (I_{lq})_i  =\left\{
      \begin{array}{l}
  1 \;\;\; \mbox{if} \;\;  (V_{l})_i = V_{lq} \cr
   0 \;\;\; \mbox{elsewhere}
              \end{array}            
                         \right.
      \end{displaymath}
$V_l$ can be replaced by $\{ I_{lq},\;\;  1\leq q \leq d_l \}$.
 The variable of interaction
  obtained from them is $V_k : V_l$ with $d_l$-dimension and  can be replaced by $\{ V_k:I_{lq},\;\;  1\leq q \leq d_l \}$.
  Each $V_k:I_{lq}$ is defined as :
   \begin{displaymath}     
   (V_k:I_{lq})_i  =\left\{
      \begin{array}{l}
  (V_k)_i \;\;\; \mbox{if} \;\;  (I_{lq})_i = 1 \cr
   0 \;\;\; \;\;\;\;\;\;\mbox{elsewhere}
              \end{array}            
                         \right.
      \end{displaymath}
  \item \textbf{Non-numerical variable crossed with  non-numerical variable} :\\
   $V_k$   and $V_l$ are two  non-numerical variables with $d_k$ and $d_l$ modalities respectively.
 The variable of interaction obtained from  $V_k$   and $V_l$ is $V_k : V_l$. $V_k : V_l$ is  $d_k \times d_l$-dimension, 
  can be replaced by $\{I_{kp} : I_{lq},\;\; 1 \leq p \leq d_k\;\; \mbox{and} \;\;1 \leq q \leq d_l \} $. Each $I_{kp} : I_{lq}$ is  defined as :
  \begin{displaymath}     
   (I_{kp} : I_{lq})_i  =\left\{
      \begin{array}{l}
  1 \;\;\;\;\;\;\mbox{if}\;\; (I_{kp})_i  =  (I_{lq})_i =1  \cr
   0 \;\;\;\;\;\;\mbox{elsewhere}
              \end{array}            
                         \right.
      \end{displaymath}
 
  \item \textbf{ Identifiability of variables} :\\
  For the identifiability of variables including those of interactions, a vector $\mathcal{H}$ of integer is automatically generated,
  $\mathcal{H}=\{ h_1, h_2, \ldots, h_{N_{cov}}\}$. If $\mathcal{V}$ is the set of all covariables including interactions  then 
  $\mathcal{V}=\{ V_1, V_2, \ldots, V_{N_{cov}}\}$. $\mathcal{H}$ and $\mathcal{V}$ are two vectors with the same  length $N_{cov}$. 
  The component $h_s$   
  of $\mathcal{H}$ is the dimension of the covariable $V_s$, $1 \leq s \leq N_{cov} $.
  In the process of variables selection, even if  a non-numerical
  variable  $V_s$ is replaced by the indicators of its modalities, the indicators are automatically identified and grouped according 
  to the component $h_s$ of $\mathcal{H}$ 
  corresponding to this covariable.
 \end{enumerate}


\subsection*{Methods}
The cohort studies generally generate a high data base containing dozens of variables. In the process of  analysis of this data, 
the experts in medicine and epidemiology, use their empirical knowledge 
on phenomenon to perform pre-treatments which consists in recoding some variables and in 
choosing some interactions based on expertise.
They use classical variables selection methods like wrapper (forward, backward, stepwise, etc.), 
embedded,  filter and ranking to perform variable selection \cite{Guyon03anintroduction}. The goal of  wrapper method
is to select  subset of variables with a lower prediction error. 
The wrapper algorithm is improved by structural 
wrapper to obtain a sequence of nested subset of features for optimality \cite{DBLP:conf/esann/Bontempi05}. 
In  practice,  classical methods of features selection are practically impossible in high dimension
because  the number of features subsets given by  ($2^p$), where $p$ is the number of features,  increases.\\
The statistical analysis was conducted in three steps.
First, the  variables selection is performed using GLM-lasso method through a cross validation with two levels. At the second step,
the 
selected  variables are debiased by a GLM and used to predict the number of anopheles. At the last step, the  results are compared
to those of  reference method to clarify which of both  methods of variables selection and prediction is better.
\subsubsection*{Model}
The statistical analysis is based on GLM and data are processed using the Lasso method.  
Such  approach 
is called. GLM-Lasso \cite{kouwayejds, Kouwayemldm}. The target variable, the number of anopheles conditionally
follows a Poisson law. The Poisson laws constitute an exponential family of dispersion
and the function density of probability is :
\begin{eqnarray}
\mathcal{P}(y| \mu) &= &e^{-\mu}\frac{\mu^{y}}{y!}\cr
\mathcal{P}(y| \mu) &=&\frac{1}{y!}\exp\{y\theta-e^{\theta}\}
\end{eqnarray}
with $\theta = \log(\mu)$. Its unity variance function is $\mu$ and the deviance associated is  defined as :
\begin{eqnarray}
d(y| \mu) &=& -2\int_{y}^{\mu} \frac{y-u}{u} du\cr
d(y| \mu)&=&-2\left\{ (y-y\log(y))-(\mu-y\log(\mu))    \right\}
\end{eqnarray}
This function is convex, its  minimum is null and  obtained at $\mu = y$. This implies that $d(y| \mu)$ is positive.
The function density of probability can be defined using the deviance as : 
\begin{equation}
\label{probability_deviance}
\mathcal{P}(y| \mu) =\frac{y^y e^{-y}}{y!}\exp \left\{-\frac{1}{2}d(y|\mu) \right\}
\end{equation}
According to the  Equation \ref{probability_deviance}, minimizing the deviance is equivalent to maximizing 
the likelihood.
For each observation $i$, the  Equation \ref{probability_deviance} is defined as:
\begin{equation}
\mathcal{P}(y_i|\mu(x_i,\beta)) =\frac{y_i^{y_i} e^{-y_i}}{y_i!}\exp \left\{-\frac{1}{2}d(y_i|\mu(x_i,\beta) \right\}
\end{equation}
A simple GLM model under matrix shape is :
 \begin{equation}
g[E(Y|X,\beta)]=X\beta 
\end{equation}
where the distribution of $Y$ conditional to $(X=x)$
is the  Poisson distribution of parameter  $E(Y |X=x, \beta)$,
 $ X$ is the $n \times \,(p+1)$-dimension matrix of 
covariables (environmental variables), $n$\, is the number of observations, $p$ is the number of
covariables, 
$\beta$ is a $(p+1)$-vector of fixed parameters  including the intercept, 
$Y$ is the vector  of  the target variable.
\begin{equation}
 (Y=y_i|X=x_i) \sim \mathbb{P}(\mu_i); \;\;  
\end{equation}
 where  \;\; $ x_i\beta=\log(\mu_i)$ and $ \mathbb{P}(\mu_i)$ is a Poisson distribution of parameter $\mu_i$.
Then 
\begin{equation}
 \mathbb{P} (Y=y_i|X=x_i) = \frac{e^{(x_i\beta)^{y_i}}}{(y_i)!} \times e^{-e^{x_i\beta}}
\end{equation}
where $y_i$ is an integer, $x_i$  a vector $(x_{i1}, \ldots, x_{ip})$ of real numbers. If $\mathcal{D} = \{(Y=y_i,X=x_i), 1 \leq i \leq n\}$, the likelihood on $n$ observations can be defined as
 \begin{equation}
  L_{GLM}(\beta|\;\mathcal{D}) = \prod_ {i=1}^{n} \frac{e^{(x_i\beta)^{y_i}}}{(y_i)!} \times e^{-e^{x_i\beta}}
 \end{equation}
  and the log-likelihood is
\begin{eqnarray}
  \mathcal{L}_{GLM}(\beta|\;\mathcal{D}) &= &\log\left( \prod_ {i=1}^{n} \frac{e^{(x_i\beta)^{y_i}}}{(y_i)!} \times e^{-e^{x_i\beta}}\right)\cr
  \mathcal{L}_{GLM}(\beta|\;\mathcal{D})  &=& -\sum^{n}_ {i=1}\log((y_i)!)+  \sum^{n}_ {i=1} y_i(x_i\beta) - e^{(x_i\beta)}
 \end{eqnarray}
 Minimizing the deviance  under the constraint $\sum_ {i}|\beta_j|<t$ which is equivalent to $\lambda \sum_ {j}|\beta_j|<1$ is reduced
to minimizing without constraint on the vector $\beta$ of parameters of the regression function $ \sum_i d(y_i|\mu(x_i,\beta)+\lambda \sum_ {j}|\beta_j|$
\begin{eqnarray}
 \sum_i d(y_i|\mu(x_i,\beta)+\lambda \sum_ {j}|\beta_j|&=&-2\sum_j(y_i-y_i\log(y_i))-(\mu(x_i,\beta)-y\log(\mu(x_i,\beta))\cr
 & &+\lambda \sum_ {j}|\beta_j|\cr
 &=&+2(\sum_j (\mu(x_i,\beta)-y\log(\mu(x_i,\beta)) + \frac{1}{2}\lambda \sum_ {j}|\beta_j|)\cr
 & &-2\sum_j(y_i-y_i\log(y_i))
\end{eqnarray}
The quantity $\sum_j(y_i-y_i\log(y_i))$ does not 
 depend on  the parameter  $\mu$ of the model then minimizing $ \sum_i d(y_i|\mu(x_i,\beta)+\lambda \sum_ {j}|\beta_j|$
is reduced to minimizing 
 $(\sum_j (\mu(x_i,\beta)-y\log(\mu(x_i,\beta)) + \frac{1}{2}\lambda \sum_ {j}|\beta_j|)$. Using $\lambda$ at the place of
$\frac{1}{2}\lambda$, we can minimize 
$(\sum_j (\mu(x_i,\beta)-y\log(\mu(x_i,\beta)) +\lambda \sum_ {j}|\beta_j|)$. 
If
\begin{equation}
Q= \sum_j (\mu(x_i,\beta)-y\log(\mu(x_i,\beta)) +\lambda \sum_ {j}|\beta_j|
\end{equation}
then
\begin{eqnarray}
 Q&=&-\left(\sum_j (-\mu(x_i,\beta)+y\log(\mu(x_i,\beta)) -\lambda \sum_ {j}|\beta_j|\right)\cr
 &= &-\left( \sum_j y\log(\mu(x_i,\beta)) -\mu(x_i,\beta) + \log((y_i)!)- \lambda \sum_ {j}|\beta_j|) \right)+\sum_j\log((y_i)!)\cr
 & &   \cr
Q &=& -(L_{GLM}(\beta|\;\mathcal{D}) -\lambda \sum_ {j}|\beta_j|)+\sum_j\log((y_i)!
\end{eqnarray}
Minimizing the quantity  Q is the same thing to maximize
$L_{GLM}(\beta|\;\mathcal{D}) -\lambda \sum_ {j}|\beta_j|$. Then 
\begin{eqnarray}
\label{Log_lik_hood_pen}
\mathcal{L}_{pen}(\beta(\lambda)|\;\mathcal{D}) &=&L_{GLM}(\beta|\;\mathcal{D}) -\lambda \sum_ {j}|\beta_j|\cr
 \mathcal{L}_{pen}(\beta(\lambda)|\;\mathcal{D}) &=&-\sum^{n}_ {i=1}\log((y_i)!)  + \sum^{n}_ {i=1} y_i(x_i\beta) - e^{(x_i\beta)}-\lambda \sum_ {j}|\beta_j|
 \end{eqnarray} 
According to Equation \ref{Log_lik_hood_pen}, GLM-Lasso  is a regularizing method consisting in 
  penalizing the likelihood of the GLM by adding a penalty term 
\begin{equation}
 \mathcal{P}(\lambda) =\lambda  \sum_{i=1}^{p} {\lvert\beta_{i}\lvert}, \;\;\; \mbox{with}\;\; \lambda \geq 0
\end{equation}
\begin{eqnarray}
 \mathcal{L}_{pen}(\beta(\lambda)|\;\mathcal{D}) &=&-\sum^{n}_ {i=1}\log((y_i)!)  + \sum^{n}_ {i=1} y_i(x_i\beta) - e^{(x_i\beta)}-\mathcal{P}( \lambda)\cr
 \mathcal{L}_{pen}(\beta(\lambda)|\;\mathcal{D}) &=&L_{GLM}(\beta|\;\mathcal{D}) -\mathcal{P}(\lambda)
 \end{eqnarray} 
 The coefficients of GLM-Lasso are given by :
\begin{equation}\label{beta_lasso}
 \hat{\beta}(\lambda) = Arg\max_{\beta}{\left[L_{GLM}(\beta|\;\mathcal{D})- \mathcal{P}( \lambda)\right]}
\end{equation}
The choice of the regularizing parameter lambda is done by minimizing a score function. In  practice, 
this equation doesn't have 
 a good   numerical  solution. We can  use the combination of Laplace approximation, the 
Newton-Raphson method or  Fisher scoring to solve this problem. Such   procedure 
 is used at each learning step.
The deviance can be defined as :
\begin{equation}
 Deviance(\beta |\mathcal{D})  = \sum_{i=1}^{n}d(y_i|\mu(x_i,\beta))
\end{equation}
where \begin{equation}
       \frac{1}{2}d(y_i|\mu(x_i,\beta)=(y_i\log(y_i)-y_i)-(y_i\log(\mu(x_i|\beta))-\mu(x_i|\beta))
      \end{equation}
and $d(y_i|\mu(x_i,\beta)$ is the contribution of the observation $(y_i,x_i)$ to the deviance.
Then     
      \begin{eqnarray}
       \frac{1}{2}\sum_i d(y_i|\mu(x_i,\beta)&=&\sum_i(y_i\log(y_i)-y_i)-(y_i\log(\mu(x_i|\beta))-\mu(x_i|\beta))\cr
       &=& \sum_i(y_i\log(y_i)-y_i-\log(y_i!))\cr
       & &-\sum_i(y_i\log(\mu(x_i|\beta))-\mu(x_i|\beta)-\log(y_i!))\cr
      \frac{1}{2}\sum_i d(y_i|\mu(x_i,\beta) &=&\mathcal{L}(\mathcal{M}{(sat)}) -  \mathcal{L}(\mathcal{M}{(\beta)})\cr
     Deviance(\mathcal{M}(\beta))&=&2\left( \mathcal{L}(\mathcal{M}{(sat)}) -  \mathcal{L}(\mathcal{M}{(\beta)}) \right)
      \end{eqnarray}
where $\mathcal{M}{(sat)}$ is the ''saturated'' model and $\mathcal{M}{(\beta)}$ is the model of Poisson regression.
It is clear that $ Deviance(\mathcal{M}(Sat))=0 $.

The deviance of the  'Null' model noted $\mathcal{M}{(Null)}$ (the model with only the intercept)  is defined as : 

\begin{equation}
 \sum_{i=1}^n (y_i\log(y_i) - y_i) - (y_i\log(\bar{y})-\bar{y})
\end{equation}
then 
\begin{eqnarray}
\label{ratio_dev}
    Deviance(\mathcal{M}(\beta))&=&   2\left( \mathcal{L}(\mathcal{M}{(sat)})-\mathcal{L}(\mathcal{M}{(Null)})
    +\mathcal{L}(\mathcal{M}{(Null)}) - 
    \mathcal{L}(\mathcal{M}{(\beta)})\right)\cr
 Deviance(\mathcal{M}{(\beta)})&=& Deviance(\mathcal{M}{(Null)})-2(L(\mathcal{M}{(\beta)})-L(\mathcal{M}{(Null)}))\cr
    Deviance(\mathcal{M}(\beta))&=&Deviance(\mathcal{M}{(Null)})-ResidDev(\mathcal{M}{(\beta)})\cr
Deviance(\mathcal{M}{(\beta)})&=&Deviance(\mathcal{M}{(Null)})\left(1- \frac{ResidDev(\mathcal{M}{(\beta)})}
{Deviance(\mathcal{M}{(Null)})}\right)
\end{eqnarray}
Then we have: 
\begin{equation}
\label{dev_ratio}
 \frac{Deviance(\mathcal{M}{(\beta)})}{Deviance(\mathcal{M}{(Null)})}= 1- \frac{Deviance\; Residual(\mathcal{M}{(\beta)})}{Deviance(\mathcal{M}{(Null)})}
\end{equation}

where  $ResidDev (\mathcal{M}{(\beta)}) = 2(L(\mathcal{M}{(\beta)})-L(\mathcal{M}{(Null)})) $ is the residual deviance and 
 $ \frac{Deviance(\mathcal{M}{(\beta)})}{Deviance(\mathcal{M}{(Null)})}$ is the ratio of deviances. It is the 
  proportion of the deviance of the Null model explained by the model $\mathcal{M}{(\beta)})$. The residual deviance is positive if 
\begin{equation}\label{beta_lasso1}
 \hat{\beta}(\lambda) = Arg\max_{\beta}{\left[L(\mathcal{M}{(\beta)})- \lambda \sum_ {j=1}^p|\beta_j|\right]}
\end{equation}
suppose that : 
\begin{equation*}
 R=\frac{Deviance(\mathcal{M}{(\beta)})}{Deviance(\mathcal{M}{(Null)})}\; \mbox{and} \; r=\frac{ResidDev(\mathcal{M}{(\beta)})}{Deviance(\mathcal{M}{(Null)})}
\end{equation*}
The Equation \ref{dev_ratio} becomes :
\begin{equation}
\label{ratios}
 R=1-r
\end{equation}

Minimizing the deviance according to each value of the parameter $\lambda$ of penalty, leads to one model of parameters 
$\hat{ \beta}(\lambda)$ noted $\mathcal{M}{(\hat{ \beta}(\lambda))}$.

The aim of GLM-Lasso is to provide a model 
minimizing  the ratio $R$  
or maximizing the ratio $r$.
Equation \ref{ratios} gives 
\begin{equation}
 R= \frac{Deviance( \mathcal{M}{(\hat{ \beta}(\lambda_k))})}{Deviance( (\hat{\beta}(\lambda_{max}))} =1-r
\end{equation}
then 
\begin{equation}
\label{deviance}
Deviance(\mathcal{M}{((\hat{ \beta}(\lambda_k))}) = (1-r)\times Deviance( \mathcal{M}{( (\hat{\beta}(\lambda_{max}))})
\end{equation}
The optimal value   $\lambda.min$ of  $\lambda$  which 
 minimizes the  $Deviance$ function is:
\begin{equation}
\label{lambda.min}
 \lambda.min= Arg\min_{\lambda_{k}} [ Deviance( \mathcal{M}((\hat{\beta}(\lambda_k)))]   
\end{equation}
The value $\lambda.1se$ of $\lambda$ defined by T. Hastie et $al$  minimizes 
 the deviance plus 
its standard deviation \cite{Friedman2015glmnet, Friedman2010regularization, hastie2009element}:
\begin{equation}
\label{lambda.1se}
 \lambda.1se = Arg\min_{\lambda_{k}} [ Deviance( \mathcal{M}(\hat{\beta}(\lambda_k))) + Std( Deviance( \mathcal{M}(\hat{\beta}(\lambda_k))))].  
\end{equation}
\subsubsection*{Algorithm (LOLO-DCV)}
The algorithm Leave One Level Out Double Cross-Validation  (LOLO-DCV) developed in this work
is a  stratified cross validation    with  two  levels \cite{kouwayejds, Kouwayemldm}. The second 
level allows  
 to avoid  over-fitting in learning stage
in the process of variable selection because the number of observations is lower. 
Its aim is to compute a  second cross validation ($CV_{2}$)  for  prediction at each 
step of learning  of a first cross validation ($CV_{1}$). The predictors obtained with 
($CV_{2}$) are consistent for prediction on the test set for ($CV_{1}$).
This algorithm runs 
 as described in Algorithm ~\ref{LOLO_DCV_Algorithme}.
 \begin{algorithm}{}
 \caption{LOLO-DCV}
 \begin{enumerate}
  \item 
The data are separated  in  $N_f$-folds
 \item At each  first level $k$
 \begin{enumerate}
\item The folds are regrouped in two parts : $A_k$ and $E_k$, 
$A_k$ : the learning set 
containing the observations of $(N_f-1)$-folds, $E_k$ : the test set, 
containing the observations of the last fold. 
\item Holding-out $E_k$
\item \label{CV1} The second level of cross validation
\begin{enumerate}
\item    A full  cross validation is computed  on $A_k$ 
  \item  The two  regularizing parameters  $\lambda.min_k$ and $\lambda.1se_k$ are obtained.
  \item The coefficients  of  active variables i.e variables with non-zero coefficients 
  associated to these two parameters  are debiased
    \item Predictions are performed using a GLM model on $E_k$
  \item The presence $\mathcal{P}(X_i)$ of each variable is determined using $\lambda.min_k$ and $\lambda.1se_k$ on $A_k$
  \end{enumerate}
\end{enumerate}
\item The step ~\ref{CV1} is repeated  until predictions are performed for all observations.
\end{enumerate}
\label{LOLO_DCV_Algorithme}
\end{algorithm}

\subsubsection*{Quality criteria}

The comparison criteria used in this study are : 
\begin{enumerate}
\item The mean of predictions
\item The quadratic risk of predictions
\item The absolute risk of predictions
\item The deviance of the model
\end{enumerate}


\subsubsection*{Frequent variables}
Let  $\mathcal{V}=\{ V_1, V_2, \ldots, V_{N_{cov}}\}$  be the set of all variables including interactions.
According to the algorithm LOLO-DCV \ref{LOLO_DCV_Algorithme}, at each first level $k, 1 \leq k \leq N_f$, the second level 
of cross validation provides 
two values of lambda : $\lambda.min_{k}$ and $\lambda.1se_{k}$ Equation. \ref{lambda.min} and \ref{lambda.1se}.  $\lambda.min_{k}$ 
and $\lambda.1se_{k}$
generate two vectors $\beta(\lambda.min_{k})$ and $\beta(\lambda.1se_{k})$ of coefficients of covariables. Based on this,
one can determine the presence or the absence of each covariable.
For any  $\lambda$, let  define the function "Presence" of variable like:
\begin{displaymath}     
     \left\{
      \begin{array}{l}
  \mathcal{P}_k(V_s)  = 1 \;\;\;\mbox{if}\; \;\; \beta_s(\lambda) \neq \Theta\; \cr
  \mathcal{P}_k(V_s)  = 0 \;\;\; \mbox{elsewhere}
              \end{array}
                         \right.
      \end{displaymath}
      where $\beta_s(\lambda), \; 1 \leq s \leq N_{cov}$ is a vector of coefficients of covariables $V_s$ and $\Theta$ the null
      vector. The length of $\beta_s(\lambda)$ is function of the component $h_s$ of $\mathcal{H}$.
      For a threshold $w, \;\; 1\leq w \leq 100$, the set of frequent variables (FV) is 
      \begin{equation}
      \label{frequent_var}
       \mbox{FV}(\lambda) = \left\{ V_s, \frac{100}{N_f}\times \sum_{k=1}^{N_f}\mathcal{P}_k(V_s) \geq w\right\}
      \end{equation}
  \subsubsection*{Variables selection strategies}
Four strategies of variables selection are implemented and compared to the reference method.
The first strategy, LDLM is based on LOLO-DCV using $\lambda.min$ of Equation. \ref{lambda.min}.
The second strategy,  LDLS is based on LOLO-DCV using $\lambda.1se$ of Equation. \ref{lambda.1se}.
The third strategy FVM is based on LDLM, 
 and the last FVS is based on LDLS. At the end of the process, 
LDLM and LDLS select a best subset of covariables and these variables are used to make prediction.
The difference between these two strategies is the value of the parameter lambda in Equations. \ref{lambda.min} and \ref{lambda.1se}. For 
the third and the last strategies, at the end of  each of the first  level of the  double cross validation in LDLM and LDLS,
the presence of each
covariable is computed;  at the end of the process, a percentage of presence is evaluated. In these strategies, the minimum $s$ in Equation.
\ref{frequent_var} is 
 fixed at 100. If the presence percentage is equal or greater than the fixed minimum, this covariable is considered as present and 
 can belong to the subset of frequent variables.
The corresponding  subset  obtained by FVM and FVS  is used to predict.
 \section*{Results}
 \subsection*{Optimal subset of variables for prediction}
In  Figure. \ref{selection_variables_originales_sans_village_graph}, each vertical band  represents a variable
and the height of the band  is the frequency of the presence of the variable in the strategies FVM and FVS.
 This figure shows the results in the  Group 1. Among 136 variables, FVM selects 13 variables and FVS, 2 variables.
The selection in the  Group 2 shows that among 153 variables, FVM selects 9 variables and FVS, 2 variables.
In the Group 3, based on 136 covariables, FVM selects 11 covariables and FVS, 1 covariable. The selection in the Group 4,
shows that among 153 covariables, FVM selects 2 and FVS selects only 1 covariable.\\
\textbf{  Figure 1. Frequent variables among original variables.}
At the x-axis, are the variables including interactions and at the y-axis, the percentage of the presence of   variables.
 \subsection*{Summary of results on prediction accuracy and quality criteria}
The Table. \ref{Summary_orig} contains the results of the reference method B-GLM.
The quality of  prediction obtained  with the subset of variables  by FVM and FVS and the predictions of LDLM and LDLS on each Group
are described in Table. \ref{selection_variables_originales_sans_village_table}.
For these tables, each line represents the results of selection criteria for each strategy.
\begin{table}[!ht]
\caption{
\bf{Summary of B-GLM prediction}} \label{Summary_orig}
\begin{center}
\begin{tabular}{|l|r|r|r|r|}
\hline
  & Mean & Quadratic risk&Absolute risk&Deviance \\
\hline
B-GLM & 3.75 &62.29&3.88& 3173.9 \\
\hline
\end{tabular}
\end{center}
\end{table}
 
\begin{table}[!ht]
\caption{
\bf{LOLO-DCV on original variables} }
\label{selection_variables_originales_sans_village_table}
\begin{center}
\begin{tabular}{|l|r|r|r|r|r|r|}
\hline
   & Mean & Quadratic risk&Absolute risk& Deviance \\
\hline
LDLM & 3.75&72.04 &4.48& 5573.98 \\
\hline
LDLS & 3.75& 72.04&4.48 & 5573.98 \\
\hline
FVM & 3.75 &44.35&3.33 &3263.03\\
\hline
FVS & 3.74 &54.54& 3.66& 3698.18 \\
\hline
\end{tabular}
\end{center}
\end{table}
For the Group 1, Table. \ref{selection_variables_originales_sans_village_table}, 
FVS has the best mean of prediction, FVM is the best in deviance.
The results in  the Group 2 shows that
FVM has the best mean; it is also the best in deviance.
About the Group 3,  
FVS has the best mean; it is the best in deviance and 
for the Group 4,
FVM has the best mean; it is the best in deviance.
\subsection*{Optimal subset variables of prediction}
The best subset of variables selected for each group of covariables  is:
\begin{enumerate}
 \item \textbf{B-GLM}\\
 According to the results of B-GLM \cite{cottrell2012malaria}, the best subset of covariables is 
  Season (season), the number of rainy days during the three days of one survey (RainyDN),  
 mean rainfall between 2 survey (Rainfall), number of rainy days in the 10 days before the survey (RainyDN102), 
 the use of repellent (Repellent), The index of vegetation (NDVI), the interaction between season and NDVI (season:NDVI).
\item \textbf{LOLO-DCV (LDLM, LDLS, FVM, and FVS)}
\begin{enumerate}
 \item Based on  the results of Figure.
 \ref{selection_variables_originales_sans_village_graph} and the Table. \ref{selection_variables_originales_sans_village_table},
 the best covariables for the Group 1 (\textbf{original variables}) is 
 Season (season) and  interaction between  mean rainfall between 2 survey  and the number of rainy 
 days during the three days of one survey (Rainfall:RainyDN), results obtained by FVS.
 \item \label{best subset} For the Group 2 (\textbf{original variables with village as fixed effect}),
 the best subset of variables for optimal prediction is
  Season (season) and interaction between number of rainy days in the 10 days before the survey and village (RainyDN10:village), results obtained by FVS.
 \item 
  The calculations on Group 3 (\textbf{recoded variables}) show that the best subset of covariables 
is :
  Season (season) and mean rainfall between 2 survey (RainyDN10), results obtained by FVS.
 \item For the Group 4 (\textbf{recoded variables with village as fixed effect}),
 
 the best subset of covariables for optimal prediction is: Season (season) and interaction between the number of rainy days during 
 the three days of one survey and presence of work around the site (RainyDN:Works)
These results are obtained by FVM.
\end{enumerate}
\end{enumerate}

 \section*{Discussion}
\begin{figure}[!ht]
\begin{center}
\includegraphics[width=5in,height=3in]{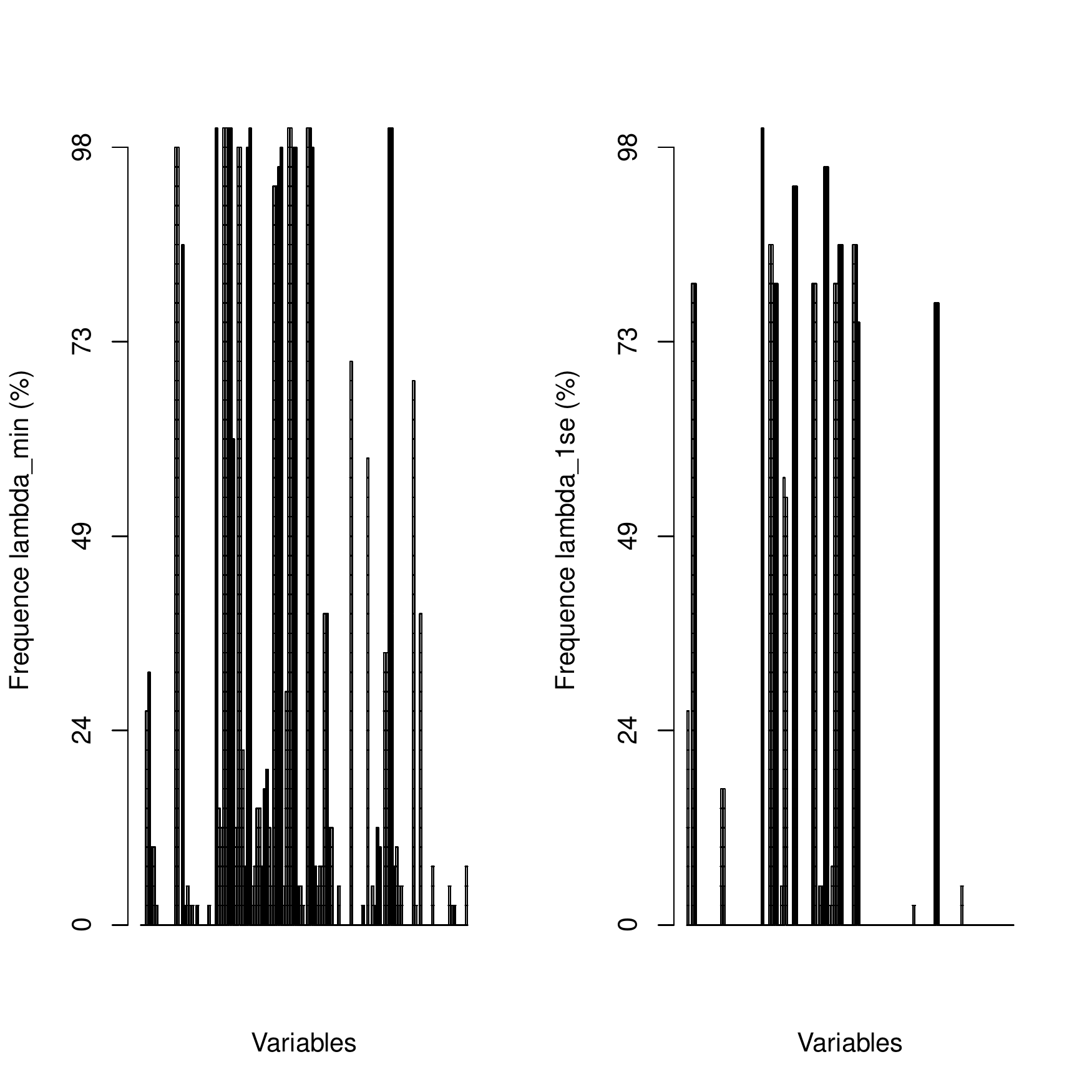}
\end{center}
\caption{
\textbf{Frequent variables among original variables.}
At the x-axis, are the variables include interactions and at the y-axis, the percentage of the presence of   variables. The sub-figure on left is 
about $\lambda.min$ and the one on right is about $\lambda.1se.$}
\label{selection_variables_originales_sans_village_graph}
\end{figure}

For each group of covariables, the best subset is selected by the trade-off between the application of
the criteria and the sparsity of the covariables subset.
Globally, the mean in prediction for the four strategies applied on the four groups of covariables is closer to the 
mean of observations 
 (3.74). 
LDLM and LDLS achieve exactly the same performance in prediction: 
mean, quadratic risk, absolute risk and  deviance.
These two strategies are approximatively 
the same  even if the subset of covariables for the optimal prediction is not the same for the different group of variables.
The mean of predictions of both   methods 
 are approximatively the same with the mean of observations (3.74) which  is achieved
exactly by FVS. In prediction, FVM and FVS  are better than LDLM and LDLS.
The algorithm LOLO-DCV shows the influence of interactions on the target variables.
The variability of the  score  in prediction
at village level (high in one the village), 
detects some problems in the data. The  Figure. 
\ref{selection_variables_originales_sans_village_graph}, shows two class of variables, the most frequent and the least frequent.
The lowest quadratic risk, absolute risk and deviance are obtained
with FVM and FVS.
FVS has the same mean in prediction with observations.
For Group 1 and Group 2, FVS is the best in prediction but for Group 3 and Group 4, FVM is the best in prediction.
The subset of covariables selected by FVS is smaller than the  one of FVM for all group of variables. The strategy 
FVS selects at most 2 variables. It is  more sparse than FVM.
FVS and FVM achieve the same performance at the absence of village Figures. \ref{selection_variables_originales_sans_village_graph}. 
The presence of village 
as variable of fixed effect strongly reduces  the number of selected variables for optimal prediction.
The number of covariables selected in Group 3 and Group 4 is lower than the one of Group 1 and Group 2. 
The number of covariables with interactions is 136 for Group 1 and Group 3 and 153 for Group 2 and Group 4.  The classical methods
will compute  $2^{136}$ or  $2^{153}$ different model  before selecting the best subset.
Combined with   double cross validation, calculation will be unrealizable  because 
of the complexity of the algorithm. The strength of LOLO-DCV is the usage of lasso and the two level cross validation. 
In a relative short time  LOLO-DCV detects all covariables selected by the B-GLM
and some interpretable  interactions among them. The Tables. \ref{Summary_orig} and  \ref{selection_variables_originales_sans_village_table}, 
show that LOLO-DCV is  the best method in selection  and  prediction.
The distribution of the prediction error according to the classes of anopheles shows
a high  variability   for B-GLM 
 and low 
  for the  LOLO-DCV. The optimal subset of features 
obtained by LOLO-DCV algorithm is  approximatively the same at each step. This proves its stability.
Finally, the best subset of variables for prediction is composed of variables selected in Group 2,
original variables with village as fixed effect, season  and interaction between the number 
of rainy days in the 10 days before the survey and village (RainyDN10 : village) \ref{best subset}. Its mean in prediction is 3.74.
 \section*{Conclusion}
In this work, we implemented an algorithm for the prediction of malaria risk  using 
environmental and climate variables. We performed the variables selection using an
automatic machine learning 
 by a method combining Lasso and stratified two levels
cross validation. The selected variables were debiased and the prediction  was achieved by simple 
GLM. The results obtained by such  procedure 
 are clearly better improved compared
to those obtained by the B-GLM method taken as the reference method. The improvement concerns all properties 
such as the quality of the selection and prediction. Moreover
, the pre-treatments 
 of experts were overcome and the CPU time used to display
our program is smaller than the one 
 required by the reference method.

%
%
%
%
%
%
\section*{Acknowledgments}
We thank all the member 
 of the laboratories:  IRD/UMR216/MERITE (Cotonou),
LERSAB (Abomey-Calavi), SAMM (Paris-France); the agencies : AUF (Agence Universitaire de la Francophonie), 
and SCAC : Service de coop\'eration et d'actions culturelles (B\'enin)
%
%
%
%
%
%
%
%
%
%
%
%
%
%

\section*{Apendix}
\textbf{ Description of original variables}
  \begin{table}[!ht]
\caption{
\bf{ Original variables}. \label{tab:Originales variables}}
\begin{center}
\begin{tabular}{|l|l|c|l|} 
\hline
\textbf{} &\textbf{Nature}&\textbf{Number of modalities}&\textbf{Modalities}\cr
\hline
Repellent&Non-numeric& 2 & Yes/ No\cr 
\hline
Bed-net&Non-numeric &2 & Yes/  No\cr
\hline
Type of roof&Non-numeric& 2 & Sheet metal/ Straw\cr
\hline
Ustensils& Non-numeric& 2 & Yes/  No\cr
\hline
Presence of constructions&Non-numeric& 2 & Yes/  No\cr
\hline
Type of soil &Non-numeric&2&Humid/ Dry\cr
\hline
Water course&Non-numeric&2&Yes/ No\cr
\hline
Majority Class&Non-numeric&3&1/4/7\cr
\hline
Season &Non-numeric&4&1/2/3/4\cr
\hline
Village&Non-numeric&9&\cr
\hline
House&Non-numeric&41&\cr
\hline
Rainy days before  mission &Numeric&Discrete&0/2/$\cdots$/9\cr
\hline
Rainy days during  mission &Numeric&Discrete&0/1/$\cdots$/3\cr
\hline
 Fragmentation Index &Numeric&Discrete&26/$\cdots$/71\cr
 \hline
Openings &Numeric&Discrete&1/$\cdots$/5\cr
\hline
Number of inhabitants &Numeric&Discrete&1/$\cdots$/8\cr
\hline
Mean rainfall &Numeric&Continue&0/$\cdots$/82\cr
\hline
Vegetation&Numeric&Continue&115.2/$\cdots$/ 159.5\cr
\hline
Total Mosquitoes &Numeric&Discrete&0/$\cdots$/481\cr
\hline
Total Anopheles &Numeric&Discrete&0/$\cdots$/87\cr
\hline
Anopheles infected&Numeric&Discrete&0/$\cdots$/9\cr
\hline
\end{tabular}
\end{center}
\end{table}
\newpage
\textbf{Description of recoded variables}
  \begin{table}[!ht]
\caption{
\textbf{Recoded variables}. Variables with star are recoded. \label{tab:Recoded variables}}
\begin{center}
\begin{tabular}{|l|l|c|l|} 
\hline
\textbf{} &\textbf{Nature}&\textbf{Number of modalities}&\textbf{Modalities}\cr
\hline
Repellent&Non-numeric& 2 & Yes/ No\cr 
\hline
Bed-net&Non-numeric &2 & Yes/  No\cr
\hline
Type of roof&Non-numeric& 2 & Sheet metal/ Straw\cr
\hline
Utensils& Non-numeric& 2 & Yes/  No\cr
\hline
Presence of constructions &Non-numeric& 2 & Yes/  No\cr
\hline
Type of soil &Non-numeric&2&Humid/ Dry\cr
\hline
Water course&Non-numeric&2&Yes/ No\cr
\hline
Majority class $^*$&Non-numeric&3&1/2/3\cr
\hline
Season &Non-numeric&4&1/2/3/4\cr
\hline
Village$^*$&Non-numeric&9&\cr
\hline
House $^*$&Non-numeric&41&\cr
\hline
Rainy days before  mission $^*$&Non-numeric&3&Quartile\cr
\hline
Rainy days during  mission &Numeric&Discrete&0/1/$\cdots$/3\cr
\hline
Fragmentation index $^*$&Non-numeric&4&Quartile\cr
\hline
Openings$^*$&Non-numeric&4&Quartile\cr
\hline
Nber of inhabitants $^*$&Non-numeric&3&Quartile\cr
\hline
Mean rainfall $^*$ &Non-numeric&4& Quartile\cr
\hline
Vegetation$^*$&Non-numeric&4&Quartile\cr
\hline
Total Mosquitoes &Numeric&Discrete&0/$\cdots$/481\cr
\hline
Total Anopheles&Numeric&Discrete&0/$\cdots$/87\cr
\hline
Anopheles infected &Numeric&Discrete&0/$\cdots$/9\cr
\hline
\end{tabular}
\end{center}
\end{table}
%
\bibliographystyle{plos2009_1.bst}
 \bibliography{Bibliography-Article-PlosOne}
\end{document}